# The Preliminary Results on Super Robustness

Qinghuai Gao

**Abstract:** In this paper, we investigate super robust estimation approaches, which generate a reliable estimation even when the noise observations are more than half in an experiment. The following preliminary research results on super robustness are presented: (1) It is proved that statistically, the maximum likelihood location estimator of exponential power distribution (or $L^p$ location estimator, for short) is strict super robust, for a given $p < 1$. (2) For a given experiment and a super robust estimator family, there is an estimator which generates an estimation that is close enough to a perfect estimation, for general transformation groups. (3) $L^p$ estimator family is a super robust estimator family. (4) For a given experiment, $L^p$ estimator on translation, scaling and rotation generates perfect estimation when $p$ is small enough, even for very noisy experiments.

1. **Parameter Estimation Problem**

A system that transforms an input $I$ to an output $O$ with a transformation $T$ is defined mathematically as below:

$$O = T(I) \tag{1}$$

Estimating the transformation $T$ based on a group of system input and system output pairs is a central and challenging problem in many pattern matching and computer vision systems. Typical examples are medical image registration, fingerprint matching, and camera model estimation.

The following concepts will be used in this paper:

**Estimator** An estimation approach to generate the system parameters based on groups of observations.

**Experiment** A group of observations that is used to generate an estimated transformation.

**Strict robustness** The capability of an estimator that gives the perfect estimation even when the good observations are perfect but the noise observations have any possible distribution.

**Super robustness** The characteristics that an estimator generates an estimated transformation whose error to the perfect estimation is bounded even when the noise observations are majority, and they move to infinite.

Suppose $I_1, I_2 \cdots, I_N$ are $N$ inputs of a system defined in the formulae (1), where $I_i$ is a point in an Euclidean space, and $O_1, O_2 \cdots, O_N$ are the corresponding outputs, where $O_i$ is a point in an Euclidean space that may have different dimension than the input space. For a transformation $T$, we define the difference of $O_i$ and $T(I_i)$ as

$$d(O_i, T(I_i)) \tag{2}$$

Thus, the overall difference between the observed outputs and the estimated outputs based on $T$ is

$$\sum_{i=1}^{N} d(O_i, T(I_i)) \tag{3}$$

The problem to estimate $T$ becomes that find a $T_b$, which satisfies:

$$D_{T_b} = \min_{T} \sum_{i=1}^{N} d(O_i, T(I_i)) \tag{4}$$

The minimum takes on any possible transformation $T$ in a predefined transformation group. We will discuss translation at first. When $d$ is the Euclidean distance, it is the least square estimation.

With translation as the investigation target, we are dealing with the problem that estimates the transformation of a special linear system from the noisy input and output, which is different to compressive sensing [7-8] where $L^p$ is also used to generate a high resolution signal from a low resolution sampling modeled with a linear system in which the output and transformation matrix are accurate.

In this paper, we use $L^p$ to define the difference of $O_i$ and $T(I_i)$:

$$|O_i - T(I_i)|^p \qquad (5)$$

This leads to the maximum likelihood estimator of the exponential power distribution. We simply call it $L^p$ estimator.

Now, the estimation problem is converted to that find a $T_b$ that satisfies:

$$\sum_{i=1}^{N} |O_i - T_b(I_i)|^p = \min_T \sum_{i=1}^{N} |O_i - T(I_i)|^p \qquad (6)$$

To observe the robustness characteristics of $L^p$ estimator, we divide the observations into two groups: all observations in the first group are perfect observations: for $i = 1, 2, \cdots, n$, $O_i = T_b(I_i)$, where $T_b$ is the ideal transformation; all observations in the second group are noisy, that is, $O_i \neq T_b(I_i)$, where $i = n+1, n+2, \cdots, N$. In this paper, we denote the number of noise observations as $M$: $M = N - n$.

We will investigate that how much percent of the ideal output out of the total observations still results an ideal estimation $T_b$ or a reliable estimation of $T$ when $L^p$ estimator is used.

To pursue strict robustness, we expect that $n$ satisfies

$$D_T \geq D_{T_b} \qquad (7)$$

for any possible transformation $T$ in a transformation group. Or

$$\sum_{i=1}^{n} |O_i - T(I_i)|^p + \sum_{i=1}^{M} |O_{i+n} - T(I_{i+n})|^p \geq \sum_{i=1}^{M} |O_{i+n} - T_b(I_{i+n})|^p \qquad (8)$$

since $O_i = T_b(I_i)$ for $i = 1, 2, \cdots, n$.

## 2. Strict Robustness on Translation

A translation is defined as:

$$T(I) = I + a_T \qquad (9)$$

For the ideal transformation $T_b$, we define it as:

$$T_b(I) = I + a_{T_b} \qquad (10)$$

For the *i*-th observation in the first group, the difference between the observation and the output of the system with a transform $T$ is:

$$O_i - T(I_i) = T_b(I_i) - T(I_i) = a_{T_b} - a_T \qquad (11)$$

Denoting that $d_T = |a_{T_b} - a_T|$, we have:

$$n d_T^p + \sum_{i=1}^{M} |O_{i+n} - T(I_{i+n})|^p \geq \sum_{i=1}^{M} |O_{i+n} - T_b(I_{i+n})|^p \qquad (12)$$

Denote that $d_i = |O_{i+n} - T_b(I_{i+n})|$. The right side of the above inequality is

$$\sum_{i=1}^{M} d_i^p \qquad (13)$$

The left side is no less than:

$$n d_T^p + \sum_{i=1}^{M} |d_i - d_T|^p \qquad (14)$$

Thus, if

$$n d_T^p + \sum_{i=1}^{M} |d_i - d_T|^p \geq \sum_{i=1}^{M} d_i^p \quad \text{or} \quad n d_T^p \geq \sum_{i=1}^{M} d_i^p - \sum_{i=1}^{M} |d_i - d_T|^p \qquad (15)$$

then $D_T \geq D_{T_b}$.

When $d_T \leq d_i$, we have

$$d_i^p - |d_i - d_T|^p \leq d_T^p \tag{16}$$

This is because $(a+b)^p < a^p + b^p$ when $p < 1$, $a > 0$ and $b > 0$.

When $d_T > d_i$,

$$d_i^p - |d_T - d_i|^p < d_T^p \tag{17}$$

Thus, when $n \geq M$, we always have $D_T \geq D_{T_b}$ for any translation.

**Theorem 1:** $L^p$ ($p < 1$) location estimator is strict robust.

## 3. Strict Super Robustness on Translation

To make estimation simpler, without loss generality, we assume that:

$$d_1 < d_2 < \cdots < d_M \tag{18}$$

When $d_T \in [d_k, d_{k+1})$, we divide the items into two groups:

$$\sum_{i=1}^{k}\left[d_i^p - (d_T - d_i)^p\right] + \sum_{i=k+1}^{M}\left[d_i^p - (d_i - d_T)^p\right] \tag{19}$$

The first group with that $d_i$ is no larger than $d_T$ is named as TFG. The second group with $d_i > d_T$ is named as TSG. We will separately estimate the upper bounds of them.

### 3.1 $d_i$ has a uniform distribution

Before sorted, $d_i$ has a uniform distribution. After sorted, the normalized $d_i$ has a beta distribution $B(i, M+1-i)$ [3], which has a distribution function of $C_{M+1}^i x^i (1-x)^{M-i}$.

The normalized $d_{M/2}$ has a distribution function $C_{M+1}^{M/2} x^{M/2} (1-x)^{M/2}$. Based on Hoeffding's inequality [4] with $n = M+1$, $p = 1/2 + M^{a-1}$, and $k = M/2$, we have $I_{1/2 - M^{a-1}}(M/2+1, M/2+1) \leq e^{-2M^{2a-1}}$ so

$$P\left(|d_{M/2} - 1/2| < M^{a-1}\right) > 1 - e^{-2M^{2a-1}} \tag{20}$$

Similarly, the formulae above should be valid for all $d_i$. Then we have:

$$P\left(\bigcap_{i=1}^{M} |d_i - i/M| < M^{a-1}\right) > \left(1 - e^{-2M^{2a-1}}\right)^M \tag{21}$$

When $a > 1/2$, the right side goes to 1 when $M$ goes to infinite. Since $2e^{-2M^{2a-1}}$ rapidly reduces to 0, for relatively large $M$, we find an $a$ such that the right side is no less than any given percentage. The minimum $a$-s for the right side is no less than 99.9% for $M$ from 100 to 1000 are listed in the table below:

| M | a |
|---|---|
| 100 | 0.696 |
| 200 | 0.676 |
| 300 | 0.666 |
| 400 | 0.660 |
| 500 | 0.655 |
| 600 | 0.652 |
| 700 | 0.649 |
| 800 | 0.647 |
| 900 | 0.645 |
| 1000 | 0.643 |

**Table 1**. $a$ for $d_i$ close to the mean $k/M$

When $|d_i - k/M| < M^{a-1}$ for all these distances, for TFG, when $k < 2M^a$, the upper bound is $2M^a$; when $k \geq 2M^a$, the upper bound is:

$$\sum_{i=1}^{k}\left[d_i^p - (d_T - d_i)^p\right]/d_T^p = \sum_{i=1}^{k}\left[\left(\frac{d_i}{d_T}\right)^p - \left(1 - \frac{d_i}{d_T}\right)^p\right]$$

$$\leq \sum_{i=1}^{k-M^a}\left[\left(\frac{(i+M^a)/M}{k/M}\right)^p - \left(1 - \frac{(i+M^a)/M}{k/M}\right)^p\right] + M^a$$

$$\leq \sum_{i=1}^{k-M^a}\left(\frac{i+M^a}{k}\right)^p - \sum_{i=0}^{k-M^a}\left(\frac{i}{k}\right)^p + M^a \qquad (22)$$

$$= \sum_{i=k-M^a+1}^{k}\left(\frac{i}{k}\right)^p - \sum_{i=0}^{M^a}\left(\frac{i}{k}\right)^p + M^a$$

$$< 2M^a$$

So the upper bound for TFG is $2M^a$.

For TSG, we have:

$$\sum_{i=k+1}^{M}\left[d_i^p - (d_i - d_T)^p\right]/d_T^p = \sum_{i=k+1}^{M}\left[\left(\frac{d_i}{d_T}\right)^p - \left(\frac{d_i}{d_T} - 1\right)^p\right]$$

$$\leq \sum_{i=k+M^a}^{M}\left[\left(\frac{i - M^a}{d_T}\right)^p - \left(\frac{i - M^a}{d_T} - 1\right)^p\right] + M^a$$

(because $x^p - (x-1)$ is decreasing) $\qquad (23)$

$$\leq \int_{k+1}^{M+1-M^a}\left[\left(\frac{x}{d_T}\right)^p - \left(\frac{x}{d_T} - 1\right)^p\right]dx + M^a$$

$$= \frac{1}{(p+1)d_T^p}\left((M+1-M^a)^{p+1} - (M+1-M^a - d_T)^{p+1} - (k+1)^{p+1}\right) + M^a$$

$$\leq \frac{1}{(p+1)d_T^p}\left((M+1-M^a)^{p+1} - (M+1-M^a - d_T)^{p+1} - d_T^{p+1}\right) + M^a$$

Because

$$-(M+1-M^a - d_T)^{p+1}$$

$$= -\sum_{i=0}^{\infty}\binom{p+1}{i}(M+1-M^a)^{p+1}\left(-d_T/(M+1-M^a)\right)^i \qquad (24)$$

$$= -(M+1-M^a)^{p+1} + (p+1)(M+1-M^a)^p d_T + \text{many negative items}$$

, TSG is no larger than:

$$(M+1-M^a)^p d_T^{1-p} - d_T/(1+p) + M^a \qquad (25)$$

When $d_T = (1-p^2)^{1/p}M$, it has a maximum of

$$p(1-p^2)^{(1-p)/p}(M+1-M^a) + M^a \qquad (26)$$

**Theorem 2:** When $d_i$ has a uniform distribution, statistically, $L^p(p<1)$ location estimator is strict super robust for large $M$ and small $p$.

The table below shows the super robustness numerically:

| $p$ | $n/M$ |
|---|---|
| 0.50 | 0.60 |
| 0.45 | 0.57 |
| 0.40 | 0.54 |
| 0.35 | 0.51 |
| 0.30 | 0.48 |
| 0.25 | 0.44 |
| 0.20 | 0.41 |
| 0.15 | 0.38 |
| 0.10 | 0.34 |
| 0.05 | 0.30 |

**Table 1.** The numerical relation of $p$ and $n/M$ when $M=1000$ and $a=0.643$.

### 3.2 Super Robustness on More General Noise Distribution

Before sorted, $d_i$ has a distribution function $f(x)$ and a cumulative distribution function $F(x)$. Based on order statistics, the distribution of $k$-th smallest value [5] is:

$$C_M^k F(x)^k (1 - F(x))^{k-1} f(x) \tag{27}$$

For this distribution, the probability that $x$ falls in the interval $[a, b]$ is:

$$\int_a^b C_M^k F(x)^k (1 - F(x))^{k-1} f(x) dx = \int_{F^{-1}(a)}^{F^{-1}(b)} C_M^k x^k (1 - x)^{k-1} dx \tag{28}$$

Thus, we have:

$$P\left(F^{-1}(k/M - M^a) < d_k < F^{-1}(k/M + M^a)\right) > 1 - 2e^{-2M^{2a-1}} / M \tag{29}$$

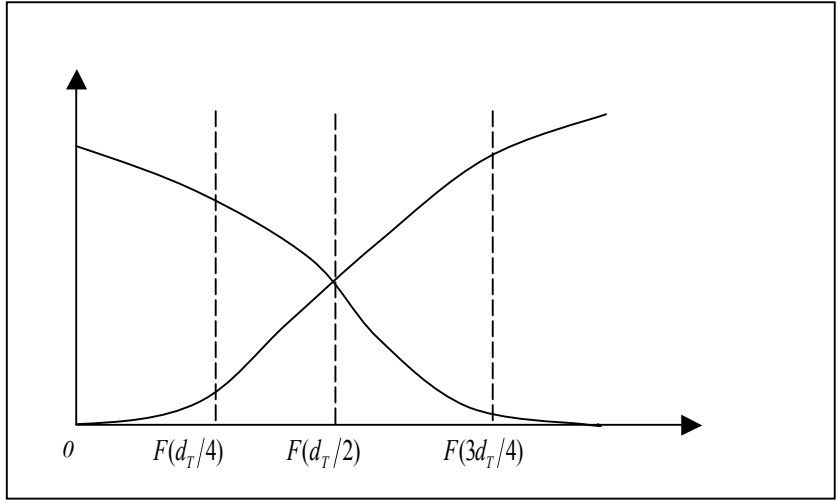

**Fig. 1** The relation of $\left(\dfrac{d_i}{d_T}\right)^p$ and $\left(1 - \dfrac{d_i}{d_T}\right)^p$

For TFG,

$$\sum_{i=1}^{k} \left[d_i^p - (d_T - d_i)^p\right] / d_T^p = \sum_{i=1}^{k} \left[\left(\dfrac{d_i}{d_T}\right)^p - \left(1 - \dfrac{d_i}{d_T}\right)^p\right] \tag{31}$$

Let's divide it into four groups as shown in Fig. 1: (1) when $d_i \leq F(d_T/4)$,

$$\frac{F^{-1}(d_i)}{d_T} \leq 1 - \frac{F^{-1}(d_i)}{d_T} - \frac{1}{2},$$

or

$$\left(\frac{F^{-1}(d_i)}{d_T}\right)^p - \left(1 - \frac{F^{-1}(d_i)}{d_T}\right)^p \leq -\frac{1}{2^p};$$

when $d_i \in [F(d_T/4), F(d_T/2)]$,

$$\frac{F^{-1}(d_i)}{d_T} \leq 1 - \frac{F^{-1}(d_i)}{d_T},$$

or

$$\left(\frac{F^{-1}(d_i)}{d_T}\right)^p \leq \left(1 - \frac{F^{-1}(d_i)}{d_T}\right)^p;$$

when $d_i \in [F(d_T/2), F(3d_T/4)]$,

$$\frac{F^{-1}(d_i)}{d_T} \leq \frac{3}{2} - \frac{F^{-1}(d_i)}{d_T},$$

or

$$\left(\frac{F^{-1}(d_i)}{d_T}\right)^p - \left(1 - \frac{F^{-1}(d_i)}{d_T}\right)^p \leq \frac{1}{2^p};$$

Thus, the overall upper bound of TFG is :

$$(F(d_T) - F(3d_T/4))M - F(d_T/4)M/2^p + (F(3d_T/4) - F(d_T/2))M/2^p. \quad (32)$$

For TSG, when $F(d_T) > 0.5$, the upper bound is $M - MF(d_T)$, in this case, we notice that the TFG upper bound is much smaller than $MF(d_T)$; when $F(d_T) \in [0.25, 0.5]$, because the function $x^p - (x-1)^p$ is decreasing, the upper bound is $((3/2)^p - (1/2)^p)(1 - F(3d_T/2))M + (F(3d_T/2) - F(d_T))M$; when $F(d_T) \in [0, 0.25]$, the upper bound is $\sum_{i=1}^{2}((i+1)^p - i^p)(F((i+1)d_T) - F(id_T))M + (1 - F(4d_T))M + (F(2d_T) - F(d_T))M$. For small $d_T$, we can obtain a lower upper bound

$$\sum_{i=1}^{\lfloor 1/d_T \rfloor - 1}((i+1)^p - i^p)(F((i+1)d_T) - F(id_T))M + (1 - F(\lfloor 1/d_T \rfloor d_T))M + (F(2d_T) - F(d_T))M$$ by repeating the same estimation approach.

Summarizing the above, we have:

**Theorem 3:** When $d_i$ has a same distribution function, statistically, $L^p (p < 1)$ location estimator is strict super robust for large $M$ and small $p$.

Specially, when all the components of the error vector has a uniform distribution, $d_i$ has a distribution of $f(x) \sim (K+1)x^K$ and $F(x)$ is $x^{K+1}$, for which, statistically, $L^p (p < 1)$ location estimator is strict super robust for large $M$ and small $p$.

## 4. Super Robustness on General Transformation Groups

### 4.1 $L^0$ Estimation

When $p = 0$, $d_0(x) = \begin{cases} 1, x > 0 \\ 0, x = 0 \end{cases}$, so we have $D_T = \sum_{i=1}^{N} d_0(O_i, T(I_i)) = N - |\{i | O_i = T(I_i)\}|$. Thus, $D_T$ reaches its minimum at $T_b$ that satisfies $|\{i | O_i = T_b(I_i)\}| = \max_T |\{i | O_i = T(I_i)\}|$. To simply the discussion below, we define:

**Precise Observation Set:** The precise observation set of a transformation instance $T$ is $\{i | O_i = T(I_i)\}$. We denote is as $POS(T)$. For $L^p (p < 1)$ estimator, we define $|POS_{p,\sigma}(T)|$ as $|\{i | d_p(O_i, T(I_i)) \leq \sigma\}|$ that shows the "support" from the experiment to the transformation.

Now we have:

**Theorem 4:** When its ideal observation group of an experiment is larger than the size of any $POS(T)$, $L^0$ estimation gives the ideal transformation $T_b$; when $|POS(T_b)| < N/2$, strict super robustness is observed.

This is a general result for the discussion in the section 3.2 of [6].

The function $f(x) \sim e^{-d_0(x)/a}$ is not a distribution function anymore. However, it will help us to establish the theory of super robustness of $L^p$ estimation as shown in the following sections.

## 4.2 Supper Robustness of $L^p$ Estimation on General Transformation

Now let's prove that there is a positive $p_0$ that $D_{T_b}^{p_0} = \min_{T, p \geq p_0} D_T^p$ or $D_{T_b}^{p_0} = \min_T D_T^{p_0}$ is good enough for a given experiment.

Suppose that the feasible solution for $T$ is a bounded compact set (this is reasonable for a given experiment).

For any given $\varepsilon$ and $A \in (0,1)$, there is a $p(\varepsilon, A)$ such that for any $p \leq p(\varepsilon, A)$, there are at most $M-1$ outputs that satisfy $|O_i - T(I_i)|^p < A$ for any $T$ that is out of the open $\varepsilon$-spheres of any perfect estimations, where $M$ is the number of ideal outputs. If not, there are infinite sequences $p_k$ and $T_k$, for which there are at least $M$ output that satisfy $|O_i - T_k(I_i)|^{p_k} < A$. Due to there are only $C_N^M$ possible combinations of these $M$ outputs, there is an index set $i_1, i_2, ..., i_M$ such that $|O_{i_j} - T_k(I_{i_j})|^{p_k} < A, j = 1,..., M$ for infinite elements of $p_k$ and $T_k$: we continue to use $p_k$ and $T_k$ to represent these two sub sequences. Since $p_k$ is a decreasing sequence with a limit 0, we have that $|O_{i_j} - T_k(I_{i_j})| < A^{1/p_k} \to 0, k \to \infty$. For the feasible solutions out of the open $\varepsilon$-spheres of any perfect estimations, split it into half with a hyper place that passes through the center and parallel to the longest side of its bounding hyper rectangular prism and choose the half that contains infinite elements of $T_k$; and continue this splitting step so that there is a sub sequence of $T_k$ which has a limit $\tilde{T}$ and $|O_{i_j} - \tilde{T}(I_{i_j})| = 0, j = 1,2,..., M$. Thus, $\tilde{T}$ is a new perfect estimation, which is a contradiction.

Now, for $p \leq p(\varepsilon, A)$, we have $\sum |O_i - T(I_i)|^p \geq (N - M + 1)A$ for estimations out of the $\varepsilon$-spheres of any perfect estimation. When $A = (N - M + 0.5)/(N - M + 1)$, we have $\sum |O_i - T(I_i)|^p \geq N - M + 0.5$. When all the perfect estimations are covered by finite number of $\varepsilon$-spheres, due to $\sum |O_i - T(I_i)|^p \to N - M, p \to 0$, there is a $p_0 \leq p(\varepsilon, A)$ for any $p \leq p_0$, at all the perfect estimations, it has $\sum |O_i - T(I_i)|^p < N - M + 0.5$. In other words, the estimation from $D_{T_b}^{p_0} = \min_T D_T^{p_0}$ is inside of the $\varepsilon$-sphere of a perfect estimation.

Because we assume that the feasible solution for $T$ is a bounded compact set, all the perfect estimations are covered by finite number of $\varepsilon$-spheres is always true.

**Theorem 5:** For any given error $\varepsilon$, there is a $p_0$ such that the estimation from $D_{T_b}^{p_0} = \min_T D_T^{p_0}$ is inside of the $\varepsilon$-sphere of a perfect estimation.

Further, for the transformation groups which satisfy $|T(I) - T_1(I)| = C(I)|T - T_1|$, we will prove that the estimation from $D_{T_b}^{p_0} = \min_T D_T^{p_0}$ is a perfect estimation for some $p_0$. To prove that, we need to prove that in a $\varepsilon$-spheres of a perfect estimation $T_b$, all the derivatives of $D_T^{p_0}$ on $|T - T_b|$ is positive. Therefore, $T_b$ reaches the local minimum. Combining with Theorem 5, $T_b$ reaches the global minimum. For a perfect observation $O = T_b(I)$, we have $|T(I) - T_b(I)| = C(I)|T - T_b|$, its contribution to $D_T^{p_0}$ is $C(I)|T - T_b|^p$. Let $y = |T - T_b|$, so $(y^p)' = py^{p-1} \to +\infty, y \to 0$. For a noise observation, its contribution to $D_T^{p_0}$ is $|O - T(I)|^p$. For a small change of $T$, the contribution change is

$$\left| |O - T(I)|^p - |O - (T+t)(I)|^p \right|$$
$$\leq p \max\left(|O - T(I)|^{p-1}, |O - (T+t)(I)|^{-1p}\right) \left| |O - T(I)| - |O - (T+t)(I)| \right|$$
$$\leq p \max\left(|O - T(I)|^{p-1}, |O - (T+t)(I)|^{p-1}\right) |T(I) - (T+t)(I)|$$
$$= p \max\left(|O - T(I)|^{p-1}, |O - (T+t)(I)|^{p-1}\right) C(I)|t|$$
$$\leq p \max\left(|(O - T_b(I))/2|^{p-1}, |(O - (T+t)(I))/2|^{p-1}\right) C(I)|t|, \quad \text{when } T \to T_b$$

. When $y = |T - T_b|$ is small enough, the overall derivative of $f(y) = \sum_{i=1}^n C(I_i) y^p + \sum_{i=1}^M \min_{y=|T-T_b|} |O_{i+n} - T(I_{i+n})|^p$ is positive. The minimum of $f(y)$ is reached at $y = 0$. Because $D_T^{p_0} \geq f(y)$ and $D_{T_b}^{p_0} = f(0)$, the estimation from $D_{T_b}^{p_0} = \min_T D_T^{p_0}$ is a perfect estimation.

**Theorem 6:** For the transformation groups which satisfy $|T(I) - T_1(I)| = C(I)|T - T_1|$, there is a $p_0$ such that the estimation from $D_{T_b}^{p_0} = \min_T D_T^{p_0}$ is a perfect estimation. Translation and uniform scaling are such transformation groups.

Further, similarly we prove that:

**Theorem 7:** For the transformation groups which satisfies $C_1(I)|T - T_1| \leq |T(I) - T_1(I)| \leq C_2(I)|T - T_1|$ for positive $C_1(I)$ and $C_2(I)$ when $|T - T_1|$ is small, there is a $p_0$ such that the estimation from $D_{T_b}^{p_0} = \min_T D_T^{p_0}$ is a perfect estimation.

Specifically, non-uniform scaling (when the components in the input vectors are all non zero) and rotation (when the origin is not an input) satisfy the condition in this theorem; therefore, most of the time, $L^p$ estimator for non-uniform scaling and rotation generates precise estimation. Also, when estimating non-uniform scaling or rotation, remove those observations whose input has zero or all zero.

**4.3 Supper Robustness of More General Estimators on General Transformation**

In the section 4.4, we only use the properties that $L^p \to L^0$ and $d^{-1}(A) \to 0$ when $p \to 0$ in the proof of Theorem 5. Now let's extend Theorem 5 to more general case.

Let's define a new family of probability distributions by distribution functions in forms of $f(x) \sim e^{-d_P(x)/a}$, where $P$ is a parameter vector to specify a distribution in the family. The difference between the observations of an experiment and the transformed results of the inputs for the maximum likelihood estimator of the distribution defined by the function $d_P$ is $D_T^P = \sum_{i=1}^N d_P(O_i, T(I_i))$.

Further, for the function family $d_P(x)$, there is a $P_0$ in the parameter space such that:

1. $d_P(x) \rightarrow L^0(x) = \begin{cases} 0, x = 0 \\ 1, x > 0 \end{cases}$, and

2. For any $A \in (0,1)$, $d_P^{-1}(A) \rightarrow 0$

, when $P \rightarrow P_0$. We name the maximum likelihood estimator of the distribution defined by the function $d_P$ with properties above as super robust estimators, or SR-estimator for short.

Similar to the proof of Theorem 5, we have that:

**Theorem 8:** For any given error $\varepsilon$, there is a $P_1$ such that the estimation from $D_{T_b}^{P_1} = \min_T D_T^{P_1}$ is inside of the $\varepsilon$-sphere of a perfect estimation.

## 5. Local Minimum Removal and Numerical Stability

The property that $d_P'(x)$ becomes infinity at each minimum helps us to prove the super robustness. However, this property generates local minima, which affects the numerical stability of the searching of the global minimum. Even though simulated annealing and genetic algorithm theoretically guarantee that the global optimum will be found, the computation cost is usually expensive. We will explore approaches to remove the local minima so that simple searching algorithm such as Simplex algorithm is good enough to find the global minimum.

We introduce the flooding estimator $d_{p,a}(x) = \begin{cases} x^p, x > a \\ x^a, x \le a \end{cases}$. The name comes from that this estimator pours "water" to fill the "pit" (the local minimum).

The global minimum from $d_{p,a}(x)$ shifts away from the minimum from $d_p(x)$. When the noise is symmetrically distributed, this shift is relatively smaller than that when the noise has a biased distribution.

It can be proved that:

**Theorem 9:** For a given experiment, there is a $A$ such that $D_T^{p,A} = \sum_{i=1}^{N} d_{p,A}(O_i, T(I_i))$ has a "bowl shape" with the condition that there is a $t_0$ such that $\frac{\partial d_{p,A}(O, T(I))}{\partial t}$ is always positive when $t \ge t_0$ and negative when $t \le -t_0$.

For a function with a bowl shape, Simplex algorithm guarantees to find the global optimum. However, if the threshold $A$ is very large, the bowl has a large flat bottom, and the search result may be meaningless.

**Theorem 10:** When $T$ is linear function of the parameters (e.g., $T(x) = x^2 t_1 - 4xt_2 + 3t_3$ with parameters $t_1$, $t_2$ and $t_3$), and there are only two minima: the global one $T_b$ and a local one $T$, the local minimum can be removed using a threshold $A$ proportional to $dis\tan ce(T_b, T) \Big/ \left[ \left( \frac{|POS_{p,\sigma}(T_b)|}{|POS_{p,\sigma}(T)|} \right)^{1/(1-p)} + C \right]$. This is also true when the local minima symmetrically distributed around the global minimum. For the translate estimator, we have $A = dis\tan ce(T_b, T) \Big/ \left[ \left( \frac{|POS_{p,\sigma}(T_b)|}{|POS_{p,\sigma}(T)|} \right)^{1/(1-p)} + 1 \right]$.

We see that when the "support" of the global minimum is much larger than that of the local minimum, and when $p$ is close to 1, the threshold $A$ is small. This means that for "good" cases, this $d_{p,a}(x)$ helps local minimum removal. For example, for translate estimator, with four local minima with the same "support" that is half of the global minimum, we have $A = dis\tan ce(T_b, T)/5$ with $p = 0.5$.

We also see that:

(1) When the noise has a biased distribution, the threshold becomes larger.

(2) There is a tradeoff between the small $p$ that generates more super robustness and a good threshold that removes more local minima.
(3) Using the threshold to decide the simplex size may help to further ignore some of the local minima.
(4) Estimating this threshold based on the experiment data, which may be a very difficult topic, is worth to investigate.

## 6. The Ebbing Algorithm

Now we introduce an ebbing algorithm to find the global minimum for general transformation:

(1) Choose a large enough $A$ ("flooding") so that $D_T^{p,A} = \sum_{i=1}^{N} d_{p,A}(O_i, T(I_i))$ has a "bowl shape".

(2) Find the global minimum (at the "bowl bottom").
(3) Gradually reduce $A$ so that the global minimum found in the last step becomes not a minimum (if it is a local minimum, shift a little) any more; further reduce $A$ more.
(4) Repeat (2) and (3) until it converges.

With this technique, we gradually avoid the local minima and find the global minima. However, when the global "bowl bottom" merges with the local "bowl bottom", the global maximum found at step (2) may be on the hillside of the local minimum of $D_T^{p,A}$ at the end of the step (3). To find the global minimum, climb to the local maximum and then use multiple paths or multiple simplexes to find the global minimum.

The name "ebbing algorithm" comes from that we gradually reduce $A$ to retreat "tide" and reveal the original global minimum.

**Theorem 11:** When $D_T^p = \sum_{i=1}^{N} d_p(O_i, T(I_i))$ only has local minima from the concaveness of $d_p(x)$ ($\frac{\partial d_p(O, T(I))}{\partial t} \to -\infty$), e. g., $T$ is linear function of the parameters, it exists a sequence of thresholds that guarantees that the ebbing algorithm finds the global minimum.

The next problem is how to find the sequence of thresholds. The following modified algorithm will resolve this problem:

(1) Choose a large enough threshold $A$ ("flooding") so that $D_T^{p,A}$ has a "bowl shape". Choose multiple points on the "bowl bottom" as start points. Reduce threshold.
(2) For each start point, search for minimum. If these minimum points locate close enough, finish searching.
(3) Check the consistency of the minima and adjust the threshold: If they are on the same "bowl bottom", they should have "same" cost function value. If majority of them has "same" cost function value, and this value is less than the values at the rest of the minimum points, choose a new group of start points on the "bowl bottom", decrease the threshold. Otherwise, increase the threshold but don't change the start points.
(4) Then, repeat (2) and (3) until the algorithm exits in (2).

It needs more complicated searching algorithms or better strategies to handle the local minima with that $\frac{\partial d_p(O, T(I))}{\partial t} = 0$ for all the parameters.

## 7. Conclusions

To summarize, these preliminary research results on super robustness are presented in this paper: (1) It is proved that statistically, $L^p$ location estimator is strict super robust, for a given $p < 1$. (2) For a given experiment and a super robust estimator family, there is an estimator which generates an estimation that is close enough to a perfect estimation, for general transformation groups. (3) $L^p$ estimator family is one of the super robust estimator families. (4) For a given experiment, $L^p$ estimator on translation and uniform scaling generates a perfect estimation when $p$ small enough, even for very noisy experiments.

The proposed approaches can be applied to many areas. Here is a short list for the areas the proposed approach dramatically changes the game: (1) estimation in very noisy experiments or environments; (2) estimation based on limited observations when find a good estimation at early stage to guide the further activities; (3) estimation when the observations are costly or lengthy; (4) analysis on the main group of the noisy observations to find the main cause of the noise. Less number of observations also means less cost or less time or both. What are estimated is not confined to physical system parameters; they can be the statistical parameters of a process: It doesn't need to wait for the sample set is large enough for the law of large number to start the estimation.